\documentclass[fleqn,10pt]{wlscirep}
\usepackage[utf8]{inputenc}
\usepackage[T1]{fontenc}
\usepackage{graphicx}
\usepackage{cite} 

\title{Acceleration noise due to Space Magnetic Field for Heliocentric Gravitational Wave Detector
}

\author[1,$\dagger$]{Peng Jia-Hui }
\author[1,$\dagger$]{Zhang Ji-Xiang  }
\author[2,$\dagger$]{Hong W. }
\author[1,*]{Su W.}
\author[3]{Ni YiWei}
\author[3]{Guo JinHan}
\author[4]{Zheng RuiSheng}

\affil[1]{MOE Key Laboratory of TianQin Mission, TianQin Research Center for Gravitational Physics \& School of Physics and Astronomy, Frontiers Science Center for TianQin, Gravitational Wave Research Center of CNSA, Sun Yat-sen University (Zhuhai Campus), Zhuhai 519082, People’s Republic of China}
\affil[2]{National Gravitation Laboratory, MOE Key Laboratory of Fundamental Physical Ouantities Measurement, and School of Physics, Huazhong University of Science and Technology, Wuhan 430074, People's Republic of China}
\affil[3]{School of Astronomy and Space Science, and Key Laboratory of Modern Astronomy and Astrophysics (Nanjing University), Ministry of Education, Nanjing University, Nanjing 210023, China.}

\affil[4]{School of Space Science and Technology, Institute of Space Sciences, Shandong University, Weihai 264209, China}

\affil[$\dagger$]{These authors contributed to the work equally.}
 
\affil[*]{suwei25@mail.sysu.edu.cn}

\begin{abstract}
The space-borne gravitational wave observatory is to detect low-frequency gravitational wave signals in the range of 0.1 mHz to 100 mHz. 
The inertial sensors of space gravitational wave require very high accuracy for acceleration noise, and the interaction of the space magnetic field with the test mass can generate magnetic moment forces and Lorentz forces, which lead to acceleration noise. Here, we obtain space magnetic field data from OMNI during 25 years from 1998 to 2022. And accordingly, we calculate the acceleration noise of space magnetic field of a heliocentric gravitational wave observatory, LISA, in more than 2 solar activity cycles. Then, we obtain the amplitude spectral densities of the acceleration noise for each day of the 25 years. We find that the median of the space magnetic field acceleration noise of LISA at 1 mHz is about $1 \times \rm 10^{-17}~m s^{-2}~Hz^{-1/2}$. We compare the space magnetic field acceleration noise of LISA and a geocentric gravitational wave observatory, TianQin, and find that the acceleration noise of the space magnetic field is of comparable magnitude for TianQin and LISA, and neither of them exceeds the respective acceleration noise requirements. Based on the statistical result of space magnetic field acceleration noise in more than 2 solar cycles, we give the $\chi$--$\xi$ parameters map of the TM for LISA and TianQin, and find that TianQin has a more stringent requirement of the parameters design than that of LISA.
\end{abstract}

\begin{document}

\flushbottom
\maketitle
%
%
\thispagestyle{empty}


\section*{Introduction}
The Laser Interferometer Gravitational-Wave Observatory (LIGO) successfully detected the first Gravitational Wave (GW) event which was generated by the merger of a binary stellar mass black holes in 2015 \cite{Abbott-2016}. So far, several other GW detectors have been established and are now operational, e.g. VirGo \cite{Acernese-2015} and KAGRA \cite{Somiya-2012}. Up to now, all the GWs event are observed by ground-based GWs detector (LIGO, VIRGO, KAGRA) with frequency range of 10 -- 1000 Hz \cite{Abbott-2019, Abbott-2021, Akutsu-2019}. Due to the impact of environments such as earthquakes and ocean current movements, ground-based GW detection is accompanied by more noise components. Additionally, due to space-scale limitations, ground-based GW detectors cannot detect GWs in the low frequency band, e.g., mHz. However, in low-frequency bands, there are abundant sources of GWs that can be used to study fundamental physics, astrophysics, and cosmology \cite{Kuroda-2015}. In order to extend the frequency range of GWs detection to mHz, some space-based GWs detection programs have gradually been proposed, such as Laser Interferometer Space Antenna (LISA) \cite{Amaro-Seoane-2017}, gLISA \cite{Tinto-2015}, which is proposed by European Space Agency (ESA) in cooperation with the United States, Taiji (TJ) \cite{XueFei-Gong-2015} and Tianqin (TQ) \cite{LuoJun-2016}, which are proposed in China, and Japan's fractional Hertz Interference Gravitational Wave Observatory (DECIGO) \cite{Kawamura-2011}. Some of these space-based GW detection projects are in heliocentric orbits, e.g., LISA, TaiJi, DECIGO, while others are in geocentric orbits, e.g., TQ.

LISA is a heliocentric GW detection project which adopts equilateral triangular constellations with arm lengths of 2.5 million kilometers \cite{Amaro-Seoane-2017}. It is planned to be deployed 
about 50 million kilometers behind the Earth, and following the Earth around the sun \cite{Amaro-Seoane-2017}. The scientific goal of LISA is to detect GW signals in the frequency range of 0.1 mHz to 1 Hz and measure the parameters of GW 
sources \cite{Amaro-Seoane-2017}. TQ is a geocentric GWs detection project, it is composed of three satellites and deployed at an altitude of 100,000 kilometers from the center of the Earth, which form an equilateral triangle with an arm length of about 170,000 kilometers \cite{LuoJun-2016}. The period of TQ satellites move around the Earth is about 3.65 days \cite{LuoJun-2016}. Both heliocentric and geocentric GWs detector will encounter space space environment issues \cite{SuWei2021,Zhiyin-Sun-2023}. Space in the vicinity of the orbit of the GWs detector is not a vacuum, there are space plasma \cite{SuWei2021}, magnetic field \cite{kianHong-Low-2024, Armano-2015}, and cosmic rays around the orbit \cite{SuWei-2020}. The space plasma can lead to laser propagation noise for LISA, TJ, and TQ, and affect the accuracy of GWs detection \cite{SuWei2021,Jennrich-2021}. The cosmic rays can penetrate the protective layer of the test mass (TM) and cause charging effect on the TM \cite{Ruilong-Han-2023}.

Magnetic field interaction with TM can generate acceleration noise \cite{SuWei2021}. Due to the presence of slight residual magnetic moment and residual charge in the TM \cite{WeiHong-Gu-2024}, this may lead to magnetic induction forces and Lorentz forces \cite{Lopez-Zaragoza-2020}. These forces are non-conservative forces, which can lead to acceleration noise. The requirement of the acceleration noise for space GWs detector is very strict, which is on the order of 10$^{-15}$ m s$^{-2}$ in the sensitive frequency band \cite{Amaro-Seoane-2017}. This means that for LISA, the influence of the space magnetic field needs to be considered \cite{Frank-2020}.
Over the past decades, LISA has study acceleration noise due to the spacecraft magnetic field \cite{Aramano-2020}. At the end of 2015, LISA PathFinder (LPF) was successfully launched, and various hardware of LISA was tested by LPF in space \cite{Armano-2018}, including the effects of space magnetic fields \cite{Aramano-2020}. LPF reported the acceleration noise caused by the space magnetic field and spacecraft magnetic field, who find that the acceleration noise in the high frequency band is mainly due to the spacecraft magnetic field, and the acceleration noise in the low frequency band is mainly due to the space magnetic field \cite{Lopez-Zaragoza-2020}.
For the other heliocentric GWs project, TJ, acceleration noise caused by the space magnetic field is also be reported in these years \cite{XingJian-Shi-2021}.
There is still a lack of research on the effect of the evolution of the space magnetic field over the entire solar activity cycle (about 11 years) on LISA.
For the geocentric GW detectors TQ, the acceleration noise caused by the space magnetic field was reported \cite{SuWei-2020}, in which the acceleration noise due to space magnetic field during a moderate solar wind conditions is evaluated based on the magnetohydrodynamic (MHD) simulation. Furthermore, the acceleration noise due to the space magnetic field during more than one solar activity cycle is estimated based on a data-driven semi-empirical space magnetic field model \cite{SuWei-2023}. Following these works, the acceleration result based on the other space magnetic field model for TQ is reported recently \cite{kianHong-Low-2024}. The evaluation of acceleration due to space magnetic field provides constraints and information on the design of the TM \cite{JiaHao-Xu-2022}, and the design and fabrication of the TM has also been progressing \cite{FangChao-Yang-2020, HangTan-Yin-2021, AoYu-Lou-2023}.

The contents of this paper are as follows: The second part introduces the calculation of the magnetic acceleration noise caused by the space magnetic field; The third part is the analysis and discussion of the results for the TM magnetic acceleration noise. The fourth part is the final summary of the article.

\section*{Results}

\subsection*{Space magnetic acceleration model}

In this work, we will focus on the effect of acceleration noise caused by space magnetic field.
In actual production and manufacturing, TM cannot be made non-magnetic, but carries a trace of magnetic moment \cite{Lopez-Zaragoza-2020}. The TM with magnetic moment will be subjected to magnetic moment force in the background magnetic field. The dominant term of magnetic moment force due to space magnetic field is as follow \cite{SuWei2021},
\begin{equation}
 a_{\mathrm{M}} = \frac{1}{m \xi_{\mathrm{m}}} [ (\frac{2 \chi_{\mathrm{m}}V_{\mathrm{m}}B_{\mathrm{sp}}}{\mu_{\mathrm{0}}}) \cdot \nabla] B_{\mathrm{sc}}
\label{equation:a_m}
\end{equation}
where, $m$ is the mass of the TM, $\xi_{\mathrm{m}}$ is the magnetic shielding factor, $\chi_{\mathrm{m}}$ is the magnetic susceptibility of the TM, $V_{\mathrm{m}}$ is the volume of TM, $B_{\mathrm{sp}}$ is the space magnetic, $B_{\mathrm{sc}}$ is the spacecraft magnetic, and the $\mu_{\mathrm{0}}$ is the vacuum permeability.

In addition, due to the action of energetic particles such as galactic cosmic rays (GCRs) and solar energetic particles (SEPs) will make the TM charged. 
The charged TM will be subjected to the Lorentz force in the background magnetic field. The Lorentz force due to space magnetic field is as follow, 
\begin{equation}
 a_{\mathrm{L}} = \frac{\eta}{m} q \boldsymbol{v} \times \boldsymbol{B}_{\mathrm{sp}}
\label{equation:a_L}
\end{equation}
Here, $\eta$ is an effective shielding factor, $q$ is the charge of the TM, $v$ is the speed of the TM.

The parameters of LISA used in this work are as follows: mass of the TM, $m$ = 1.96 kg; side length of the TM, $r$ = 4.6 cm; magnetic susceptibility of the TM, $\chi_{\mathrm{m}}$ = $2.5 \times 10^{-5} $; magnetic shielding factor $\xi_{\mathrm{m}}$ = 10, 
residual magnetic moment of the TM, $M_{\mathrm{r}}$ = $2 \times 10^{-8}$ $\mathrm{Am^{2}}$; spacecraft magnetic field, $B_{\mathrm{sc}}$= $1 \times 10^{-6}$ T; distance between the equivalent magnetic moment source in the spacecraft and TM, $r_{\mathrm{sc}}$ = 0.75 m \cite{Schumaker-2003, Diaz-Aguil-2011, Lopez-Zaragoza-2020}.
The parameters of TQ used in this work are as follows: $m$ = 2.45 kg, $r$ = 5 cm, $\chi_{\mathrm{m}}$ = $1 \times 10^{-5} $, $\xi_{\mathrm{m}}$ = 10, $M_{\mathrm{r}}$ = $2 \times 10^{-8}$ $\mathrm{Am^{2}}$, $B_{\mathrm{sc}}=1.6\times10^{-6}$ T \cite{LuoJun-2016, SuWei-2020, SuWei-2023}.

LISA is planned to launch around 2035, the initial scientific observation time is 4 years, and an additional 6 years working time if everything goes smoothly, the total running time of LISA can reach 10 years \cite{Amaro-Seoane-2017}.
The solar is the source of the space plasma and magnetic field. The solar magnetic poles reverse every 11 years, the period of the solar activity cycle is about 11 years. 
And there is a cycle with about 11 years in the evolution of interplanetary solar wind parameters, e.g. plasma number density, plasma bulk speed, and magnetic field \cite{Toledo-2023,Poopakun-2023}.
The runtime of LISA (about 10 years) is close to solar activity cycle (about 11 years), it reminds us that the acceleration due to the space magnetic field during the solar cycle needs to be considered.


In this work, we used the OMNI dataset to get the space magnetic field data. The time resolution of the space magnetic field is 60 seconds, the duration of the data are from 1998 to 2022 (25 years). With the space magnetic field data from OMNI of 25 years (more than 2 solar activity cycles), which can fully cover the 2 solar active cycles, we can get evolutionary characteristics of space magnetic acceleration throughout the whole solar cycles for LISA.
The OMNI solar wind data is mainly composed of in-situ observations, such as Advanced Composition Explorer (ACE) \cite{JH-King-2005} and WIND \cite{Ogilvie-kw-1995}. It contains physical parameters such as solar wind speed, plasma number density, interplanetary magnetic field, space weather indices such as $D$st index, Sym-H index, Kp index, and so on. 


Here we show the space magnetic field of 6 days during 1998 to 2022 in Figure \ref{fig:Bsp-t} as an example. 
The components of space magnetic field $B_\mathrm{x}$, $B_\mathrm{y}$ and $B_\mathrm{z}$ are on the order of nT. 
As shown in Figure \ref{fig:Bsp-t}, the fluctuation of space magnetic field can reach large values on the order of 10 nT (as shown in 2018-01-01), and small values on the order of nT (as shown in 2013-01-01), respectively.

\subsection*{The acceleration noise caused by space magnetic field for LISA}


\begin{figure}
\centering
\includegraphics[width = 35pc]{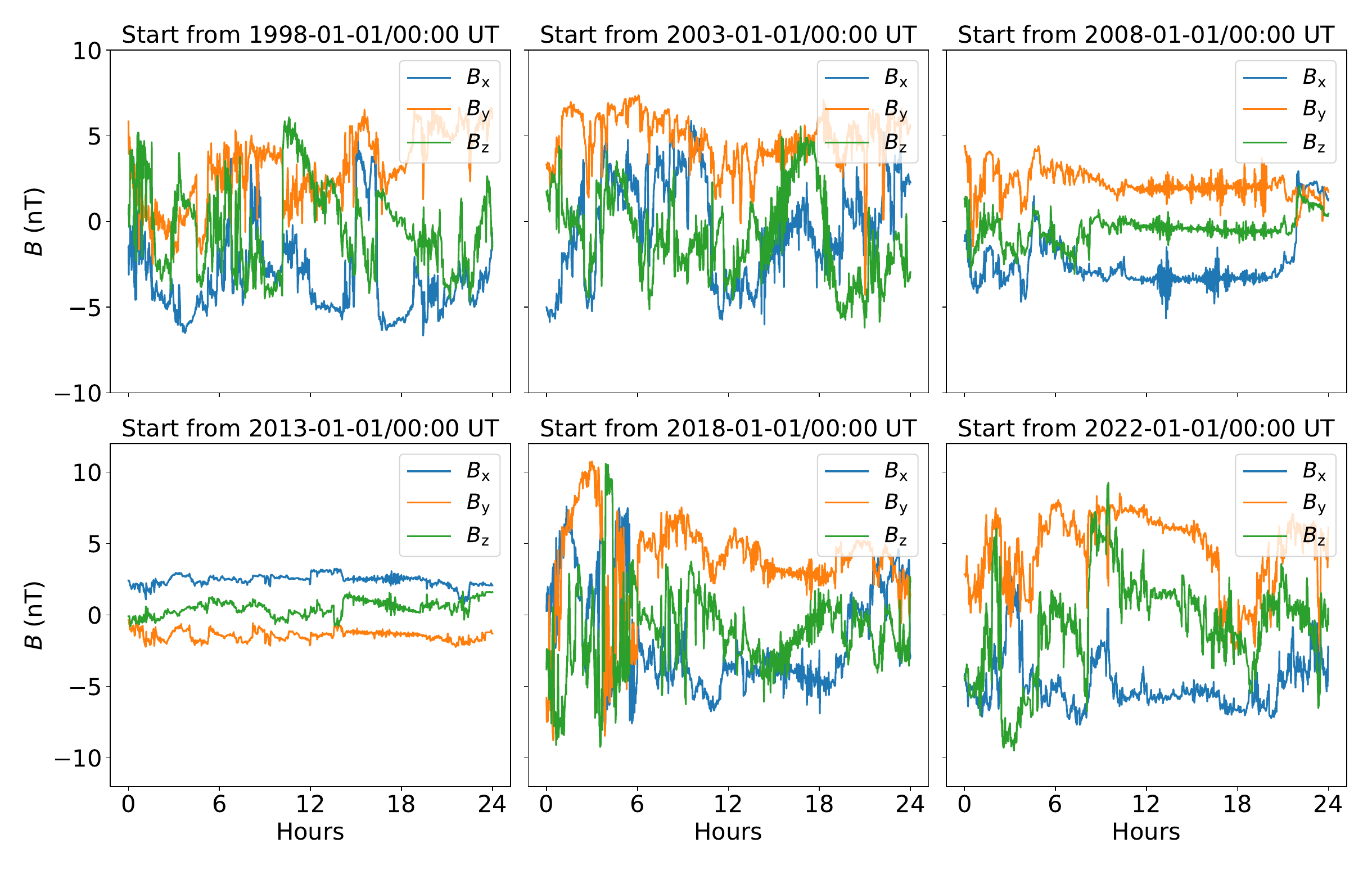}
\caption{The space magnetic field of 6 days during 1998 to 2022.}
\label{fig:Bsp-t}
\end{figure}

\begin{figure}
\centering
\includegraphics[width = 35pc]{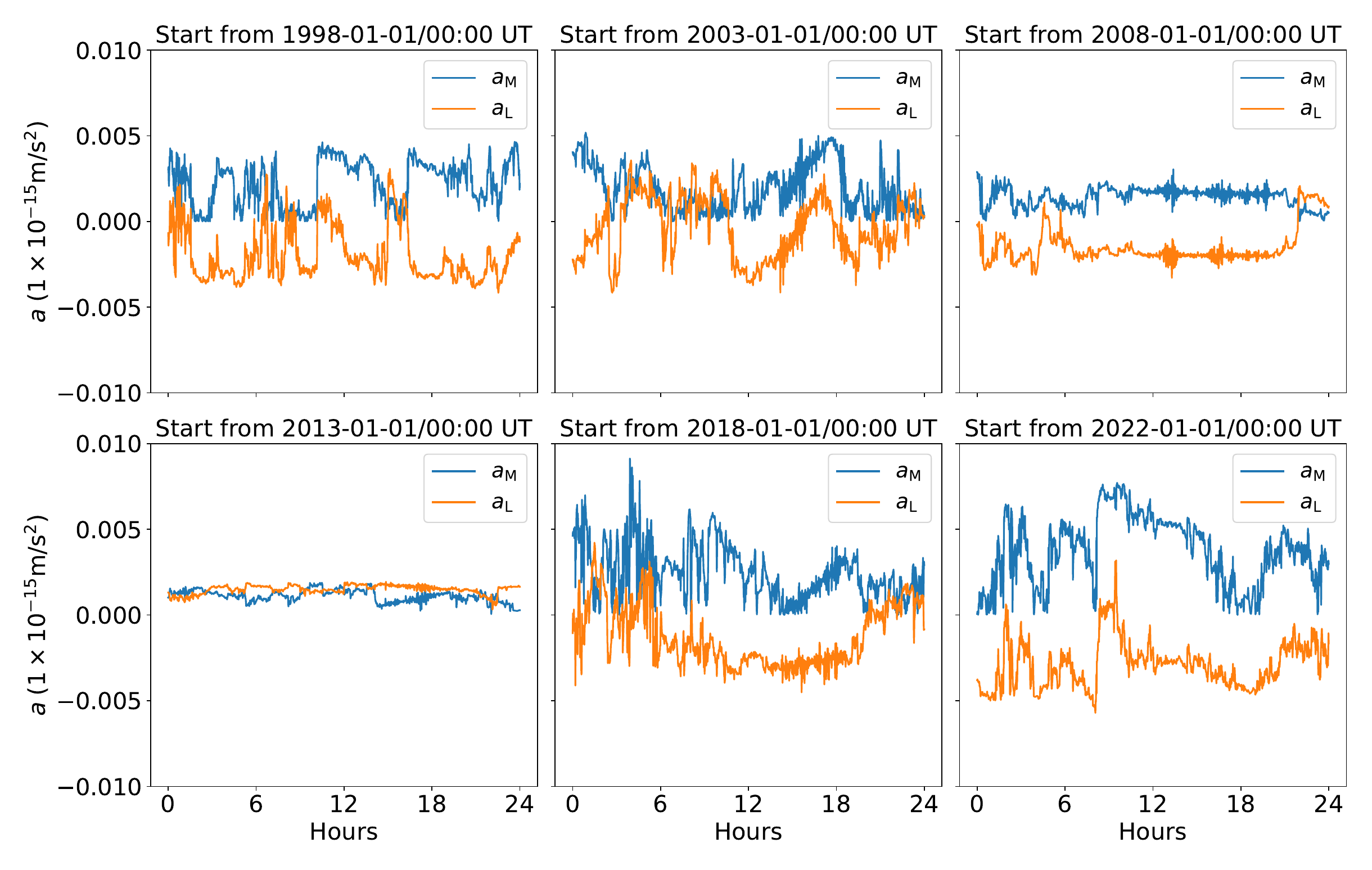}
\caption{The time series of $a_{\mathrm{M}}$ (blue) and $a_{\mathrm{L}}$ (orange) of the six days during 1998 to 2022.}
\label{fig:a-t}
\end{figure}

Combining the space magnetic field data from OMNI and Equations \eqref{equation:a_m} and \eqref{equation:a_L}, we calculate the time series of space magnetic acceleration noise $a_{\mathrm{M}}$ and $a_{\mathrm{L}}$. 
As shown in Figure \ref{fig:a-t}, they are the time series of $a_{\mathrm{M}}$ and $a_{\mathrm{L}}$ for 6 days in the range 1998 to 2022. 
$a_{\mathrm{L}}$ and $a_{\mathrm{M}}$ are both on the order of $10^{-17}$ m s$^{-2}$.
Compared with Figure \ref{fig:Bsp-t}, it shows that the larger the magnetic field, the larger the corresponding $a_{\mathrm{M}}$ and $a_{\mathrm{L}}$. It is due to the magnetic field force and Lorentz are both proportional to the magnetic field.

\begin{figure}
\centering
\includegraphics[width = 35pc]{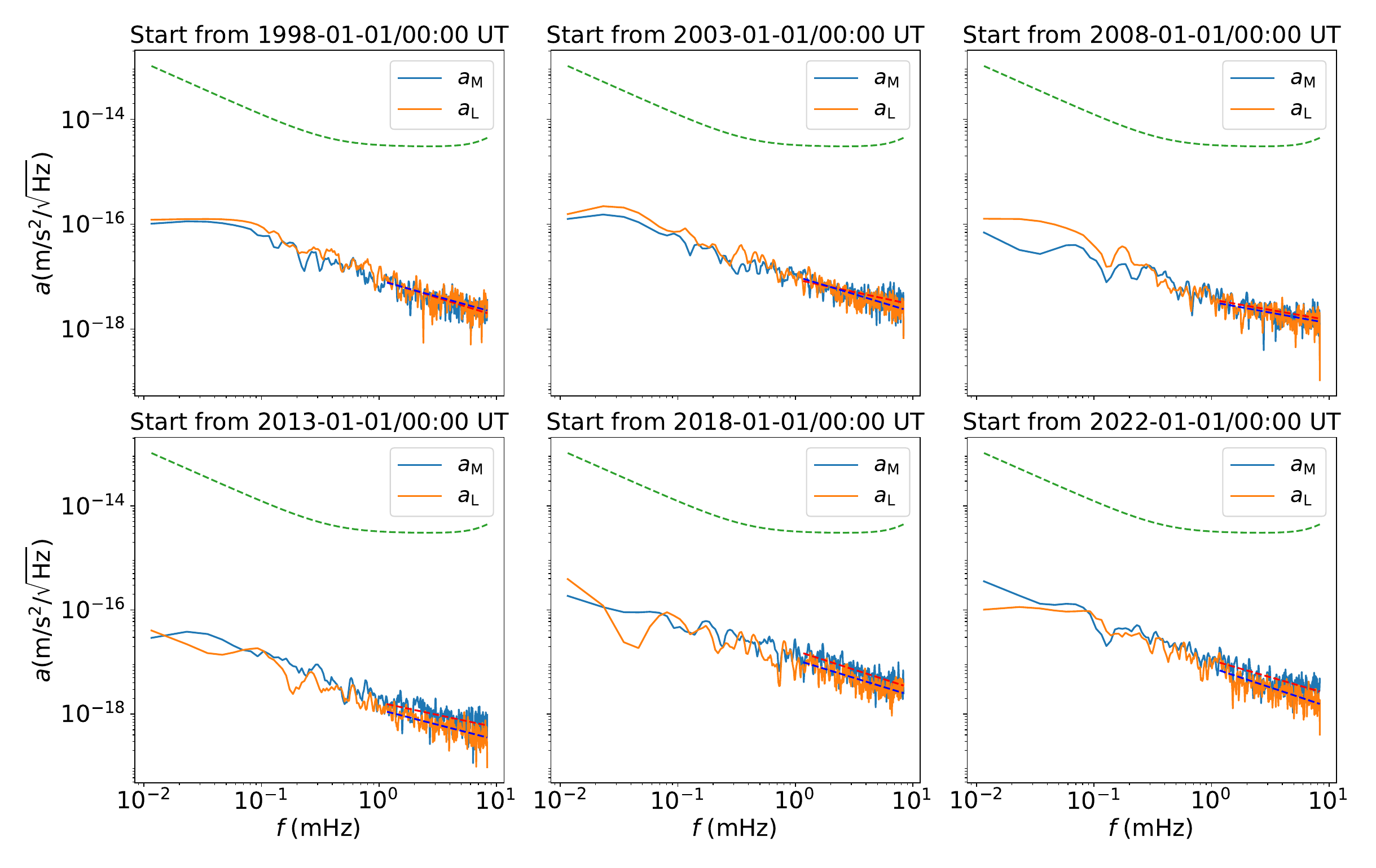}
\caption{ASDs of $a_{\mathrm{M}}$ and $a_{\mathrm{L}}$ in the six days during 1998 to 2022. The ASD of $a_{\mathrm{M}}$ and $a_{\mathrm{L}}$ are blue and orange curves, and the green dashed curve is the acceleration requirement of LISA.}
\label{fig:ASDs-f}
\end{figure}

Furthermore, we calculate the amplitude spectral density (ASD) of $a_{\mathrm{M}}$ and $a_{\mathrm{L}}$ based on the time series of acceleration noise.
We calculate the ASDs of the magnetic acceleration noise $a_{\mathrm{M}}$ and $a_{\mathrm{L}}$.
Figure \ref{fig:ASDs-f} are the ASDs of $a_{\mathrm{M}}$ and $a_{\mathrm{L}}$ of the 6 cases in Figure \ref{fig:a-t}. The solid red lines in Figure \ref{fig:ASDs-f} are the result of the fitting of the single power law after smoothing by using the Savitzky-Golay filter \cite{Savitzky-1964}. The ASDs of $a_{\mathrm{M1}}$ and $a_{\mathrm{L}}$ are larger at low frequencies and gradually decrease with increasing frequency. They are typical color noise, and for LISA, the spectra of the ASDs of $a_{\mathrm{M}}$ and $a_{\mathrm{L}}$ are about -0.695 and -0.793, respectively. And for TQ, the spectra of the ASDs of $a_{\mathrm{M}}$ and $a_{\mathrm{L}}$ are about -0.753 and -0.723, respectively.
The requirement of the acceleration noise for LISA is taken as follow \cite{Amaro-Seoane-2017},
\begin{equation}
    S_{a}^{1/2} \le 3 \times 10^{-15} \frac{\mathrm{m s^{-2}} }{\sqrt[]{\mathrm{Hz}} } \cdot \sqrt[]{1 + (\frac{0.4  \mathrm{mHz} }{f})^{2} } \cdot \sqrt[]{1 + (\frac{ f }{8 \mathrm{mHz} })^{4} }
\end{equation}
where, $f$ is frequency. The acceleration noise requirement curve of LISA is shown as green dotted lines in Figure \ref{fig:ASDs-f}.
Comparing the requirement curve with the ASDs of the acceleration noise caused by the space magnetic field, we find that the acceleration noise is about two orders of magnitude smaller than the requirement curve of LISA. 
It shows that with the given values of the parameters,
the space magnetic field has a modest effect on LISA with the existing parameter values.


\subsection*{The magnetic acceleration noise in solar cycles}

\begin{figure}
\centering
\includegraphics[width = 15 cm]{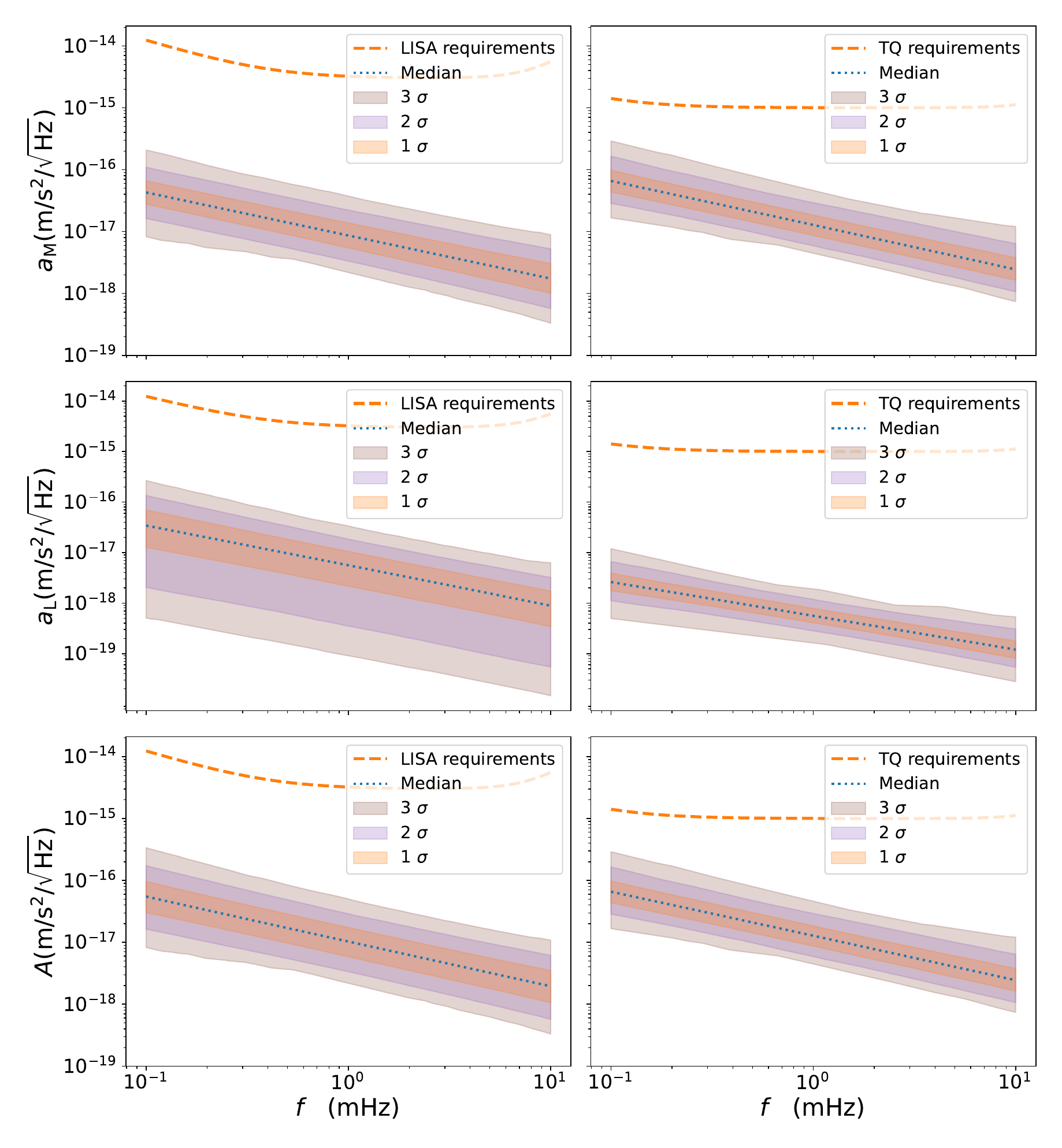}
\caption{The statistical results of space magnetic field acceleration noise ASDs for LISA and TQ. The left panels are the statistical results of $a_{\rm M}$, $a_{\rm L}$, and $A$ for LISA, and the right panels are the statistical results of $a_{\rm M}$, $a_{\rm L}$, and $A$ for TQ. The median of the acceleration noise is represented as blue line. The orange, purple, and brown shading are 1-$\sigma$, 2-$\sigma$, and 3-$\sigma$ intervals of the ASDs of the acceleration noise.
}
\label{fig:ASDs-Statistics}
\end{figure}

As the planned duration of the LISA mission is comparable to a solar cycle, we study the space magnetic acceleration during more than two solar cycles for LISA here.
We calculate the ASDs of $a_{\rm M}$ and $a_{\rm L}$ for every day from 1998-01-01 to 2022-12-31 (25 years), and the results are shown in the left panels of Figure \ref{fig:ASDs-Statistics}.
In this work, 
the ratio of the accelerations to the acceleration requirement at 0.4 mHz, 1 mHz, and 8 mHz are denoted as $R_{\rm 0.4mHz}$, $R_{\rm 1mHz}$, and $R_{\rm 8mHz}$, respectively. 
The statistical results show that $R_{\rm 0.4mHz}$, $R_{\rm 1mHz}$, and $R_{\rm 8mHz}$ of $a_{\rm M}$ for LISA are 0.00429 $\pm$ 0.00214, 0.00300 $\pm$ 0.00153, 0.000559 $\pm$ 0.000328, respectively. And $R_{\rm 0.4mHz}$, $R_{\rm 1mHz}$, and $R_{\rm 8mHz}$ of $a_{\rm L}$ for LISA are 0.00311 $\pm$ 0.00228, 0.00205 $\pm$ 0.00150, 0.000327 $\pm$ 0.000271, respectively.

In order to evaluate the total effect of the space magnetic fields, i.e. the magnetic moment force and the Lorentz force, on the detection of GWs, we define an acceleration $A = \sqrt{ a_{\mathrm{M}}^{2} + a_{\mathrm{L}}^{2} }$, to obtain the largest possible estimation of the space magnetic acceleration noise.
Subsequently, we calculated the ASDs of $A$ for every day from 1998-01-01 to 2022-12-31, and the statistical results of the 25 years are shown in the left panels of Figure \ref{fig:ASDs-Statistics}.
The orange, purple and brown shades in Figure \ref{fig:ASDs-Statistics} represent the 1-$\sigma$, 2-$\sigma$ and 3-$\sigma$ intervals, respectively. 
$R_{\mathrm{0.4mHz}}$, $R_{\mathrm{1mHz}}$ and $R_{\mathrm{8mHz}}$ of $A$ are 0.00549 $\pm$ 0.00277, 0.00374 $\pm$ 0.00193, 0.000667 $\pm$ 0.000395, respectively. 
As shown in Figure \ref{fig:ASDs-Statistics}, the statistical results of ASDs fo $A$ do not exceed LISA's requirement curve in the frequency range from 0.1 to 10 mHz, it indicates that the space magnetic acceleration can adequately meet LISA's requirements under the parameters in this work.

\begin{figure}
\centering
\includegraphics[width = 16cm]{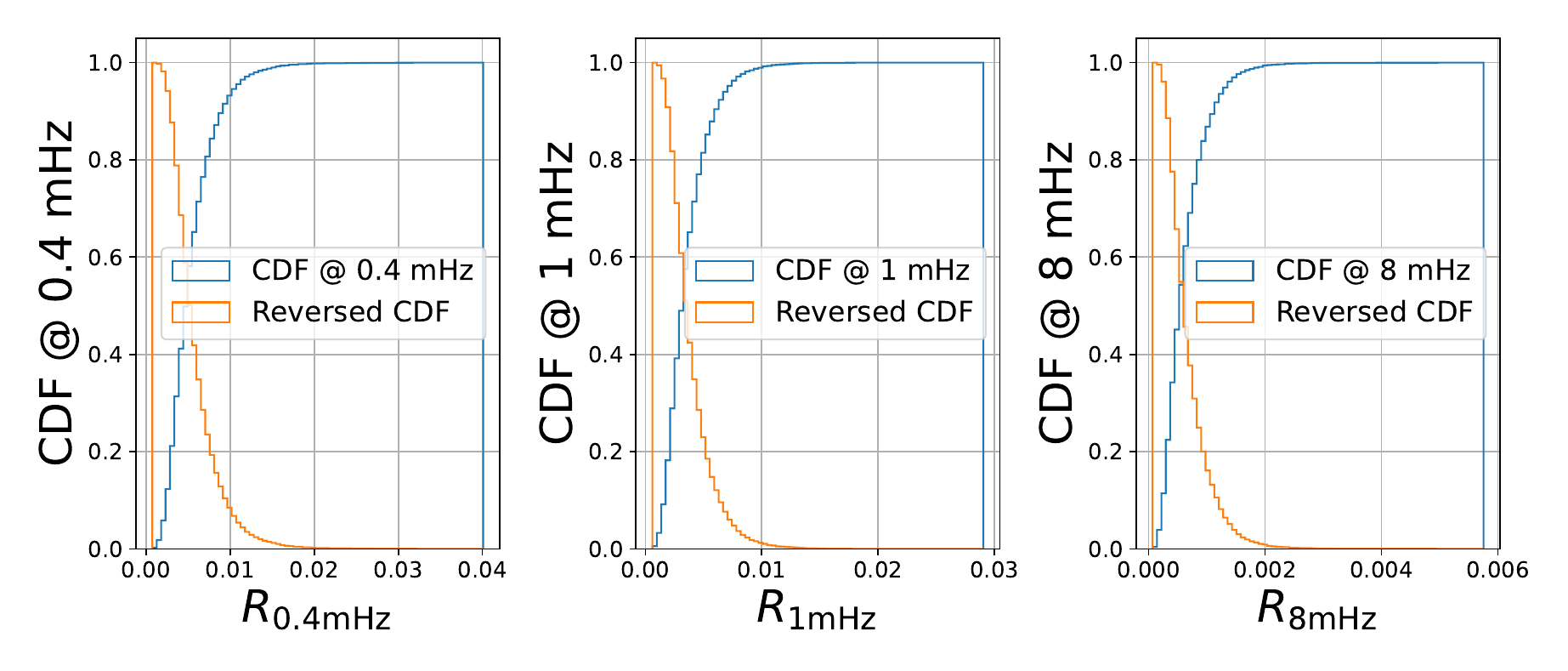}
\caption{The CDF of the magnetic acceleration noise $A$ at 0.4 mHz, 1 mHz, and 8 mHz. The blue curve and the orange curve are the CDF and the reverse CDF, respectively.}
\label{fig:CDF-A}
\end{figure}

We calculate the cumulative distribution function (CDF) of the magnetic acceleration noise $A$, and the result is shown in Figure \ref{fig:CDF-A}. The left, middle, and right panels in Figure \ref{fig:CDF-A} represent the CDFs of $R_{\mathrm{0.4mHz}}$, $R_{\mathrm{1mHz}}$, and $R_{\mathrm{8mHz}}$ for $A$, respectively. The blue bins represent the CDF, and the orange bins represent the reverse CDF. According to the reversed CDFs \ref{fig:CDF-A}, We can get the occurrence probabilities of $R_{\mathrm{0.4mHz}}$, $R_{\mathrm{1mHz}}$, and $R_{\mathrm{8mHz}}$. The occurrence probabilities of $R_{\mathrm{0.4mHz}} > 0.01$ and $> 0.02$ are 7.26\% and 0.16\%, The occurrence probabilities of $R_{\mathrm{1mHz}} > 0.01$ and $> 0.02$ are 1.23\% and 0.0219\%
not more than 2 \%. 
And the occurrence probabilities of $R_{\mathrm{8mHz}} > 0.01$ and $R_{\mathrm{8mHz}} > 0.02$ are 1.08\% and 0. 
Overall, the occurrence rate of the magnetic acceleration noise exceeding 8\% for LISA's acceleration requirement is very rare, which is less than 8\% for the sensitivity frequency range of LISA. 
It indicates that the magnetic acceleration noise caused by the interplanetary magnetic field is unlikely to exceed the requirements for LISA.

\subsection*{Comparison of magnetic acceleration noise between LISA and TQ}

Here, we update the statistical result of the space magnetic field acceleration noise of TQ, and the results are shown in the right panels of Figure \ref{fig:ASDs-Statistics}. 
The median of the amplitude of $a_{\rm M}$, $a_{\rm L}$, and $A$ at 1 mHz for LISA are $9.686 \times 10^{-18}$, $6.560 \times 10^{-18}$ and $1.208 \times 10^{-17}$ m s$^2$; The median of the amplitude of $a_{\rm M}$, $a_{\rm L}$, and $A$ at 1 mHz for TQ are $1.410 \times 10^{-17}$, $6.203 \times 10^{-19}$ and $1.412 \times 10^{-17}$ m s$^2$.
The amplitudes of $a_{\rm M}$ for LISA and TQ are of the same order. Considering that the value of $\chi_{\rm M}$ is taken to be 2.5 times that of LISA, if $\chi_{\rm M}$ of both LISA and TQ are taken to be the same, the amplitude of $a_{\rm M}$ of TQ is about 4 times that of the LISA. However, the amplitude of $a_{\rm L}$ for LISA is one order of magnitude higher than that of TQ.
For LISA, the ratio of $a_{\rm M}$, $a_{\rm L}$, and $A$ to the acceleration requirement of LISA at 1 mHz are $0.00300$, $0.00203$, and $0.00374$.
For TQ, the ratio of $a_{\rm M}$, $a_{\rm L}$, and $A$ to the acceleration requirement of TQ at 1 mHz are $0.0140$, $0.000617$, and $0.0140$.
In this work, the magnetic shielding factor $\xi_{\rm m}=10$ for both LISA and TQ, without magnetic shielding, $a_{\rm M}$ of both LISA and TQ will grow 10 times its original value.

The main reason for the difference between the results of LISA and TQ is the different space environment around their orbits. 
In general, the space magnetic field near TQ's orbits are more complex and commonly stronger than that near LISA.
For LISA, its satellites move around the Sun about 50 million kilometers behind the Earth, its orbit a is completely immersed in the solar wind.
The situation is more complicated for TQ, whose altitude determines that its orbit passes through the region where the Earth's magnetosphere and the solar wind interact, e.g. magnetopause and bow shock\cite{Wang2018,Lv2021}. But its orbit is mainly within the Earth's magnetosphere, with only a small portion of the time spent within the solar wind. 
Inside the magnetopause is the region dominated by the Earth magnetic pressure and outside the magnetopause is the region dominated by the solar wind plasma dynamic pressure. Moreover, due to compression by the bow shock of Earth, the magnetic field of the bow shock downstream (magnetosheath) is stronger than that of the bow shock upstream (solar wind). It results in the magnetic field around the TQ orbit being several times stronger than that around the LISA orbit, thus $a_\mathrm{m}$ of TQ is about several times larger than that of LISA.
The orbit of TQ is mostly inside the Earth's magnetosphere, and the velocity of the TQ satellite relative to the Earth's magnetosphere is about 2 km s$^{-1}$; Whereas the orbit of LISA is in the solar wind, and the velocity of the LISA satellite relative to the interplanetary magnetic field is about 30 km s$^{-1}$, which is about 15 times of the velocity of TQ. The Lorentz force is proportional to both the velocity and the magnetic field, which leads to the final result that Lorentz force of LISA of the 
different order of TQ, even though the magnetic field in the vicinity of LISA's orbit is weaker than that around TQ's orbit.

\begin{figure}
\centering
\includegraphics[width = 16cm]{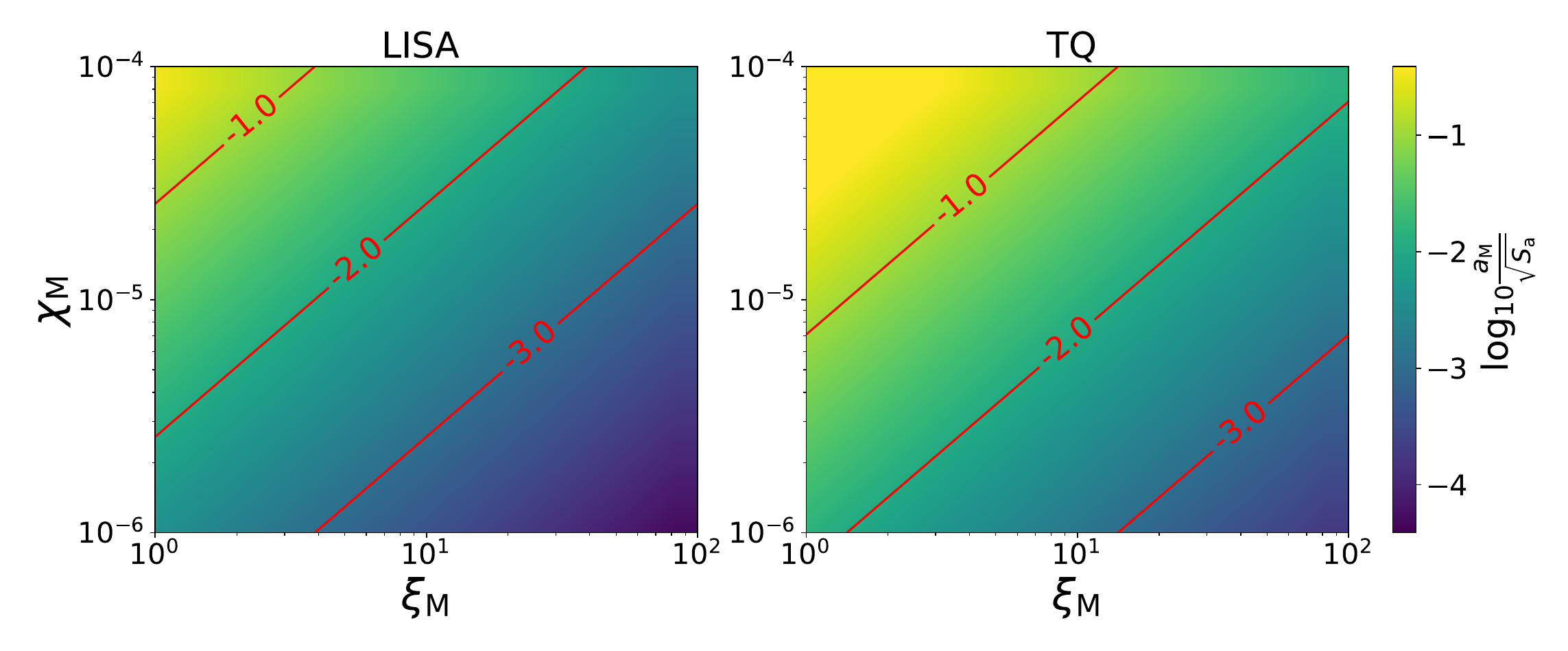}
\caption{The parameters space of $\xi_{\mathrm{m}}$--$\chi_{\mathrm{m}}$ at 1 mHz for LISA (left panels) and TQ (right panels). The color-bar represent $a_{\mathrm{M}/\sqrt{S_a} }$. The contours of  $a_{\mathrm{M}}/\sqrt{S_{\mathrm{a}}} = 10^{-1}$, $10^{-2}$, and $10^{-3}$ are marked as contours lines of -1, -2, and -3, respectively.}
\label{fig:2D-parameter}
\end{figure}

The acceleration noise due to the space magnetic field is related to the manufacturing and design level of the TM. 
The magnetic susceptibility $\chi_{\rm M}$ and the magnetic shielding factor $\xi_{\rm M}$ are the key parameters that affect the acceleration noise. As shown in Equation $\eqref{equation:a_m}$, $a_{\rm M}$ is directly proportional to $\chi_{\rm M}$ and inversely proportional $\xi_{\rm M}$. Both LISA and TQ have stringent requirements for $\chi_{\rm M}$, and both on the order of $10^{-5}$. In order to further reduce the acceleration noise caused by the magnetic field, both LISA and TQ consider adding magnetic shielding. 
Based on the statistical results of LISA and TQ for $a_{\rm M}$ in more than 2 solar cycles, we give the design parameters for $\chi_{\rm M}$ and $\xi_{\rm M}$ here. By using the median value of $a_{\rm M}$ for LISA and TQ at 1 mHz as the baselines, we estimate the design parameter space of $\chi_{\rm M}$ and $\xi_{\rm M}$ for LISA and TQ in Figure \ref{fig:2D-parameter}. The left panel of Figure \ref{fig:2D-parameter} is the LISA parameter space of $\chi_{\rm M}$--$\xi_{\rm M}$, and the right panel is $\chi_{\rm M}$--$\xi_{\rm M}$ space for TQ. The red contour lines marked with -1, -2, and -3 represent the contours where $\xi_{\rm M}$ reaches the acceleration noise requirements of $10^{-1}$, $10^{-2}$, and $10^{-3}$ for the GW detectors. 
Looking at the left boundary of $\chi_{\rm M}$--$\xi_{\rm M}$ space for TQ, the point ($\chi_{\rm M}=10^{-5}, \xi_{\rm M}=1$) is higher than the contour of $10^{-1}$. 
It indicates that, in the case of $\chi_{\rm M}$ on the order of $10^{-5}$, the magnetic shielding is necessary to ensure that $a_{\rm M}$ is less than 10\% of TQ's acceleration requirements at 1 mHz.
It can be seen that contours of TQ are lower in the $\chi_{\rm M}$--$\xi_{\rm M}$ space than that of LISA, suggesting that TQ requires more stringent parameters than LISA.


\section*{Discussion}

In this work, we obtain space magnetic field data from OMNI during 1998-01-01 to 2022-12-31 (25 years), and based on the data, we calculate the acceleration noise due to space magnetic field for LISA. We obtain the ASDs of the space magnetic acceleration noise for 25 years (longer than 2 solar cycles), and evaluate the acceleration noise due to the space magnetic field for LISA during the solar activity cycles. We find that the $a_{\rm M}$ and $a_{\rm L}$ of LISA both are on the order of 10$^{-17}$ m s$^{-1}$ when magnetic shielding $\xi_m=10$. The ASDs of the median values of $a_{\rm M}$, $a_{\rm L}$, and the total space magnetic acceleration noise $A$ are $9.686 \times 10^{-18}$ $\mathrm{m s^{-2}}$ and $6.516 \times 10^{-18}$ $\mathrm{m s^{-2}}$, $1.205 \times 10^{-17}$ $\mathrm{m s^{-2}}$ at 1 mHz, which do not exceed the acceleration noise requirement of LISA. In 0.1 mHz to 10 mHz band, the occurrence rate of space magnetic acceleration noise exceeding the LISA requirement by 1\% is less than 10\%. Due to the similar space magnetic and plasma environment, the results in this work also informative for other heliocentric GW detectors such as TJ. We update the space magnetic acceleration noise for TQ, and compare the acceleration noise due to the space magnetic field for LISA and TQ: for $a_{\rm M}$, the magnetic acceleration noise for TQ is slightly higher than that of LISA, but still in the same order; For $a_{\rm L}$, TQ's $a_{\rm L}$ is about one order of magnitude smaller than that of LISA, which is due to the orbits of LISA and TQ and their space environments in the vicinity of the orbits; For the total acceleration $A$, the results of LISA and TQ are close. Based on the statistical results of the space magnetic acceleration noise of LISA and TQ for more than 2 solar cycles, we estimate the parameters design space of $\chi$--$\xi$ for LISA and TQ. It shows that $\chi$--$\xi$ of TQ needs a more stringent requirement than that of LISA. In particular, when taking $\chi = 10^{-5}$ without magnetic shielding ($\xi = 1$), the space magnetic acceleration noise of TQ will exceed the TQ requirement by 10\%. Overall, taking the present design parameters of LISA and TQ, the ratio of space magnetic field acceleration noise to acceleration demand is higher for TQ than for LISA, but both of the acceleration noises due to space magnetic field are lower than the respective acceleration noise requirements.


\section*{Methods}


The TM is in the inertial sensors\cite{ChengLei-Yue-2024}, it is an alloy cube with 73$ \% $ gold and 27$ \%$ platinum \cite{Amaro-Seoane-2017}.
The TM has a residual magnetic moment and the magnetic susceptibility of the TM is not zero, thus, there is a magnetic force for the TM with a magnetic moment in the background magnetic field \cite{Lopez-Zaragoza-2020}.
And the magnetic force of the TM can be expressed as \cite{SuWei-2020},
\begin{equation}
 \boldsymbol{F} = \nabla(\boldsymbol{M}_{\mathrm{TM} } \cdot \boldsymbol{B} )
\label{eq:1}
\end{equation}
where, $\boldsymbol{M}_{\mathrm{TM} } $ is the magnetic moment of the TM, and $\boldsymbol{B}$ is the magnetic field. The magnetic field $\boldsymbol{B}$ is composed of the space magnetic field $B_{\mathrm{sp}}$ and the spacecraft magnetic field $B_{\mathrm{sc}}$, $\mathit{B} = B_{\mathrm{sp}} + B_{\mathrm{sc}}$. $M_{\mathrm{TM}}$ can be decomposed into residual magnetic moment $M_{\mathrm{r}}$ and induced magnetic moment $M_{\mathrm{i}}$, $M_{\mathrm{TM}} = M_{\mathrm{r}} + M_{\mathrm{i}}$. The induced magnetic moment is divided into two parts, one caused by $B_{\mathrm{sc}}$ and the other by $B_{\mathrm{sp}}$.
Thus, $M_{\mathrm{TM}}$ can be written as:
\begin{equation}
 M_{\mathrm{TM}} = M_{\mathrm{r}} + M_{\mathrm{i}} = M_{\mathrm{r}} + \frac{\chi_{\mathrm{m}}V_{\mathrm{m}}B_{\mathrm{sc}}}{\mu_{\mathrm{0}}} + \frac{\chi_{\mathrm{m}}V_{\mathrm{m}}B_{\mathrm{sp}}}{\mu_{\mathrm{0}}}
\label{eq:2}
\end{equation}
where $\chi_{\mathrm{m}} $ is the magnetic susceptibility of the TM, $V_{\mathrm{m}}$ is the volume of the TM, $m$ is the mass of the TM, $\mu_{\mathrm{0}}$ is the vacuum permeability. 
In this work, we denote $M_{\mathrm{isc}} =\chi_{\mathrm{m}}V_{\mathrm{m}}B_{\mathrm{sc}}/\mu_{\mathrm{0}}$, and $M_{\mathrm{isp}} =\chi_{\mathrm{m}}V_{\mathrm{m}}B_{\mathrm{sp}}/\mu_{\mathrm{0}}$.
Combining Equations (\ref{eq:1}) and (\ref{eq:2}),
the magnetic acceleration noise caused by the magnetic moment can be written as $a_{\mathrm{M}}$ in equation of section \ref{equation:a_m},

Since $M_{\mathrm{r}}$ is approximately stable \cite{John-Hanson-2003, YinHang-2021, Lopez-Zaragoza-2020}, $M_{\mathrm{r}}$ is taken as a constant in this work, and the surface current on the TM is ignored \cite{JiaHao-Xu-2022}. 
Since the house of the TM can provide magnetic and electric shielding \cite{Sumner-TimothyJ-2020}, the magnetic and electric shielding factors, $\xi_{\mathrm{m}}$ and $\xi_{\mathrm{e}}$ are introduced here. 
In addition, we focus on the effect of the space magnetic field $B_{\mathrm{sp}}$ here, the fluctuations of spacecraft magnetic field $B_{\rm sc}$ is neglected. 
Finally, the variation terms related to space magnetic field are as follows,
\begin{equation}
\left\{
\begin{array}{l}
a_{\mathrm{M1}} = \frac{1}{m \xi_{\mathrm{m}}} \left[\left(M_{\mathrm{r}} + \frac{2 \chi_{\mathrm{m}} V_{\mathrm{m}} B_{\mathrm{sp}}}{\mathrm{\mu_{0}}}\right) \cdot \nabla\right] B_{\mathrm{sc}} \\
a_{\mathrm{M2}} = \frac{1}{m \xi_{\mathrm{m}}} \left[\left(M_{\mathrm{r}} + \frac{2 \chi_{\mathrm{m}} V_{\mathrm{m}} B_{\mathrm{sc}}}{\mathrm{\mu_{0}}}\right) \cdot \nabla\right] B_{\mathrm{sp}} \\
a_{\mathrm{M3}} = \frac{1}{m \xi_{\mathrm{m}}} \left[\left(M_{\mathrm{r}} + \frac{2 \chi_{\mathrm{m}} V_{\mathrm{m}} B_{\mathrm{sp}}}{\mu_{0}}\right) \times \left(\frac{\mathrm{\varepsilon_{0}} \mathrm{\mu_{0}} \partial E_{\mathrm{sc}}}{\partial t}\right)\right] \\
a_{\mathrm{M4}} = \frac{1}{m \xi_{\mathrm{m}}} \left[\left(M_{\mathrm{r}} + \frac{2 \chi_{\mathrm{m}} V_{\mathrm{m}} B_{\mathrm{sc}}}{\mu_{0}}\right) \times \left(\frac{\mathrm{\varepsilon_{0}} \mathrm{\mu_{0}} \partial E_{\mathrm{sp}}}{\partial t}\right)\right] \\
a_{\mathrm{M5}} = \frac{1}{m \xi_{\mathrm{m}}} \cdot \frac{2 \chi_{\mathrm{m}} V_{\mathrm{m}} B_{\mathrm{sc}}}{\mathrm{\mu_{0}}} \nabla B_{\mathrm{sc}} \\
a_{\mathrm{M6}} = \frac{1}{m \xi_{\mathrm{m}}} \cdot \frac{2 \chi_{\mathrm{m}} V_{\mathrm{m}} B_{\mathrm{sp}}}{\mathrm{\mu_{0}}} \nabla B_{\mathrm{sp}}
\end{array}
\right.
\label{equation:4}
\end{equation}
where $\varepsilon_{\mathrm{0}}$ is the vacuum permittivity, $E_{\mathrm{sc}}$ is the electric field inside the spacecraft at the TM position, and $E_{\mathrm{sp}}$ is the space electric field. More details are in \cite{SuWei-2020, SuWei-2023}.


The space magnetic field $B_{\mathrm{sp}}$ around the LISA orbit is generally not more than 10 nT, the fluctuation of the space electric field with time $\partial E_\mathrm{sp}/\partial t$ generally does not exceed $10^{-6}$ V m$^{-1}$ s$^{-1}$ \cite{JH-King-2004, JH-King-2005}. 
The interplanetary magnetic field gradient $\nabla B_{\mathrm{sp}} \lesssim 0.01$ nT/m \cite{SuWei-2020}.
The variation of the electric field at the TM position $\frac{\partial E_{\mathrm{sc}}}{\partial t}$ is on the order of $10^{-4}$ V m$^{-1}$ Hz$^{-1/2}$ ,
and the $\nabla B_{\mathrm{sc}}$ at the TM position of is on the order of 1 $\mathrm{nT \; m^{-1} \; Hz^{-1/2}}$ at 1 mHz \cite{Stebbins-2004, Antonucci-2011}. 
In this work, we focus on the acceleration due to space magnetic field, we disregard the effect of the satellite magnetic field, thus $a_{\mathrm{M5}}$ is ignored here.
The remaining five magnetic acceleration noises $a_{\mathrm{M1}}$, $a_{\mathrm{M2}}$, $a_{\mathrm{M3}}$, $a_{\mathrm{M4}}$, $a_{\mathrm{M6}}$ are on the orders of $10^{-15} \; \mathrm{m \; s^{-2}}$, $10^{-20} \; \mathrm{m \; s^{-2}}$, $10^{-30} \; \mathrm{m \; s^{-2}}$, $10^{-32} \; \mathrm{m \; s^{-2}}$, $10^{-23} \; \mathrm{m \; s^{-2}}$. We found that the leading term of the six magnetic acceleration noises is $a_{\mathrm{M1}}$ and $a_{\mathrm{M5}}$, but in this work, we only considered $B_{\mathrm{sp}}$, so the study of the magnetic acceleration noise $a_{\mathrm{M}}$ caused by the induced magnetic moment was converted to the study of $a_{\mathrm{M1}}$.

Since the direct current (DC) term has no effect on the ASD in frequency domain, the DC term ($M_{\mathrm{r}} \cdot \nabla B_{\mathrm{sc} }$) in $a_{\rm M1}$ can be ignored, and since $a_{\rm M1}$ is the dominant term, here, we represent $a_{\rm M1}$ as $a_{\rm M}$ in Equation \eqref{equation:a_m}.
$a_{\rm M}$ is on the order of $10^{-17} \; \mathrm{m \; s^{-2}}$.


The TM is enclosed and protected by a house in the spacecraft, but the house cannot shield all energetic particles, there are rare particles with high energy can penetrate the house and hit the TM, making the TM charged \cite{Wass-2005}.
The charged TM moving in the space magnetic field is subject to the Lorentz force, which 
is a non-conservative force and also 
has an impact on GW detection. The Lorentz force for the TM with charge $q$ in the space magntiec field is show in Equation \eqref{equation:a_L},
where $v$ is the speed of the TM, the magnetic shielding effect is considered here denoted as $\xi_{\mathrm{m}}$. 
The speed of the TM is taken as the speed of LISA satellite with the value of abouy 30 km/s in the heliocentric-ecliptic (HCE) coordinate system \cite{Rubbo-2004}. $q$ of both LISA and TQ are taken as $1.6\times10^{-12}$ C here \cite{Diaz-Aguil-2011}, and $\xi_{\rm m} = 10 $ here. $a_{\mathrm{L}}$ for LISA is on the order of $10^{-17}$ m s$^2$.

\bibliography{sample}



\section*{Acknowledgements}

S.W. is supported by the National Key R\&D Program of China (No. 2020YFC2201200), NSFC (grant No. 12473060 and 122261131504).

\section*{Author contributions statement}

P.J., H.W., and Z.J. do the most of the calculation and write the paper,  S.W. proposes the idea and provides the source code, N.Y., G.J., and Z.R. provide the key discussion. All authors review the manuscript. 





\end{document}